\documentclass[aps,prl,showpacs,twocolumn,bibliography]{revtex4-2}
\usepackage{graphicx,graphics,color,epsfig} 
\usepackage{amsmath,amssymb,amsfonts} 
\usepackage{bm,times,xspace}
\usepackage{mathrsfs,latexsym}
\usepackage{verbatim}
\usepackage[colorlinks,
citecolor=blue,
linkcolor=red,
citecolor=blue,
urlcolor=blue]{hyperref}
\usepackage[mathlines]{lineno}
\usepackage{ulem}
\usepackage[thicklines]{cancel}
\usepackage{xcolor}

\usepackage{dcolumn}

\def\bs#1{\boldsymbol{#1}}

\UseRawInputEncoding
\begin{document}
\title{Collective Interference of Phonon Spin and Dipole Moment Rotation Induced Circular Dichroism}

\author{Yizhou Liu}
\thanks{Y. Liu and Y.-T. Tan contributed equally to this work.}
\author{Yu-Tao Tan}
\thanks{Y. Liu and Y.-T. Tan contributed equally to this work.}
\author{Dapeng Liu}
\author{Jie Ren}\email{Corresponding author. E-mail: xonics@tongji.edu.cn}

\affiliation{Center for Phononics and Thermal Energy Science, China-EU Joint Lab on Nanophononics, School of Physics Science and Engineering, Tongji University, 200092 Shanghai, China}

\begin{abstract}
The classical field description of ``phonon spin" relies on the invariance of a continuous elastic field under infinitesimal rotation. However, a local ``medium element" in the continuous field may contain large numbers of vibrational particles at microscopic level, like for complex lattices with many atoms in a unit cell. We find this causes the phonon spin in real materials no longer a simple sum of each atom rotation, but a collective interference of many atoms, since phonons are phase-coherent vibrational modes across unit cells. We demonstrate the collective interference phonon spin manifested as the dipole moment rotating (DMR) of charge-polarized unit cell, by deriving the infrared circular dichroism (ICD) with phonon-photon interaction in complex lattices. We compare the DMR with the local atom rotation without interference, and exemplify their distinct ICD spectrum in a chiral lattice model and two realistic chiral materials. Detectable ICD measurements are proposed in $\alpha$-quartz with Weyl phonon near $\Gamma$ point. Our study underlies the important role of collective interference and uncovers a deeper insight of phonon spin in real  materials with complex lattices.
\end{abstract}

\maketitle

{\it Introduction}. It is well-known that classical vector fields possess intrinsic spin angular momentum (spin AM, SAM) as indicated by Noether's theorem for an infinitesimal rotational invariance of the Lagrangian density. The most famous example is the electromagnetic field with left (right) handed circular polarization that corresponds to spin $-\hbar$ ($+\hbar$) state of photon. Likely, for phonons, it is not surprising that transverse vibration waves are also spin polarized as theoretically predicted more than six decades ago~\cite{1961PS,1962PS,1962PSLevine}. Recent investigations exhibit a renaissance of study of phonon spin in elastic media~\cite{ElasticSpin2018PNAS, ElasticSpin2021NC,PS2022CPL,ElasticSAMOAM2022PRL,ElasticAM2018PRB}, acoustic systems~\cite{AcousticSpin2019NSR,AcousticSpin2019PRL, AcousticSpin2020NC, AcousticSpin2020NSR, AcousticSpin2023PRAppl, Bliokh2024Oct}, and crystalline solids~\cite{ChiralPhonon2014PRL, ChiralPhonon2015PRL,KekulePS2017PRL,ChiralPhonon2018Science, Liu2019Jul, ChiralPhonon2023Nature, ChiralPhonon2023NP,Zhanglf2022NL, Tateishi2025Apr}, which not only is of fundamental interest in wave dynamics but also exhibits exotic applications in versatile functional devices such as phononic waveguide~\cite{AcousticSOI2019NC, SAWSpinWG2023PRAppl}, selective one-way phonon transport~\cite{ElasticSpin2023PRL, ElasticSpinWG2024AS}, phonon spin selectivity induced by elastic spin-orbit interactions \cite{Yang2024Nov}, and topological phononic materials~\cite{Acoustic2021JAP,ElasticSpin2022PRL,ElasticSkyrmion2023SA, Zhang2022Feb, Zhang2025Apr}.

For the description of phonon spin in solids, two different viewpoints are usually adopted: One is based on continuous elastic field theory; The other is based on discrete lattice models. However, a significant gap exists between these two different descriptions. On one hand, the continuous elastic field theory offers a natural tool to introduce the elastic spin based on the infinitesimal rotational transformation invariance of the elastic Lagrangian density \cite{ClassicalFieldTheory, PS2022CPL, WaveSpin2025CPL}. The derived elastic spin is $\bs{S} \propto \bs{u} \times \dot{\bs{u}}$ with $\bs{u}$ being the displacement field vector of a ``medium element'', a macroscopically infinitesimal part of the medium. The macroscopically infinitesimal ``medium element'' can yet contain vast number of atoms at microscopic level, with the field vector $\bs{u} = \frac{1}{N}\sum^N_i \bs{u}_i$ ($N$ refers to the number of atoms). Thus the spin $\bs{S} \propto \sum_{i,j} \bs{u}_i \times \dot{\bs{u}}_j = \sum_i \bs{u}_i \times \dot{\bs{u}}_i + \sum_{i\ne j} \bs{u}_i \times \dot{\bs{u}}_j$, in addition to the simple sum of each atom rotation, naturally contains crossing interference terms between atoms $i$ and $j$.

On the other hand, the continuous rotational symmetry, crucial for defining SAM, is broken in discrete lattice models, down to the discrete symmetries $C_n$ ($n \in \{2,3,4,6\}$). Discrete $C_n$ symmetries are preserved only at certain high-symmetry momenta of Brillouin zone (BZ) and its eigenvalues are used to define quantized pseudo AM with undetermined multiples of $n$ \cite{ChiralPhonon2015PRL, Zhang2022Feb, Zhang2025Apr}. Otherwise, the phonon AM is actually not well-defined in the discrete lattice modes. As such, people just usually took the simple summation of each atom rotation $\bs{S} \propto \sum_i \bs{u}_i \times \dot{\bs{u}}_i$ by hand as the phonon spin AM in discrete lattices and real material calculations.

However, this ad hoc and somehow arbitrary definition of phonon spin lacks both theoretical foundation and justification. First, although the definition may be applied for the particle picture, its extension to phonon is questionable, because phonon is not a particle in real space but a collective vibration mode involving many particles. Second, as a collective motion, the phonon (the mode of many particle vibrations) is phase coherent, and this phase-coherent collective vibrations must bring interference among many particles. Therefore, many crucial questions arise: what is the real physics picture of phonon spin in complex lattices? Does it contain the collective interference across particles at different sublattices? If so, what is the manifestation of the collective many-particle-interference phonon spin? Moreover, one should propose some detectable physics effect, by utilizing the interaction of phonon to the multiple physics field.

In this Letter, we uncover the collective interference picture of phonon spin in crystalline solids that can be viewed as the collective motion of many particles in the unit cell and measured by the infrared circular dichroism (ICD). Theoretically, we revisit the phonon-photon interaction under circularly polarized (CP) infrared light excitation by a quantum perturbation method. We find the transition rates of photon absorption (emission) induced phonon creation (annihilation) depends on a physical quantity associated with the rotation of unit cell's dipole moment, named dipole moment rotating (DMR), which shares the similar form as the collective interference phonon spin AM of the unit cell. We implement the calculation of DMR on a chiral model with three-fold screw rotation symmetry, to show that different charge distribution will cause different DMR profile on the dispersion relation, leading to different selection rule of light. We then implement first-principles calculations on real chiral materials tellurium(Te) and $\alpha$-quartz with chirality-induced DMR mapping on dispersion relations, which demonstrate the consistency of theoretical model and real materials. Finally, we propose a detectable ICD measurement based on DMR spectrum to demonstrate the effect of collective interference phonon spin in $\alpha$-quartz.

\begin{figure}[tbp]
\centering	
\includegraphics[width=0.48\textwidth]{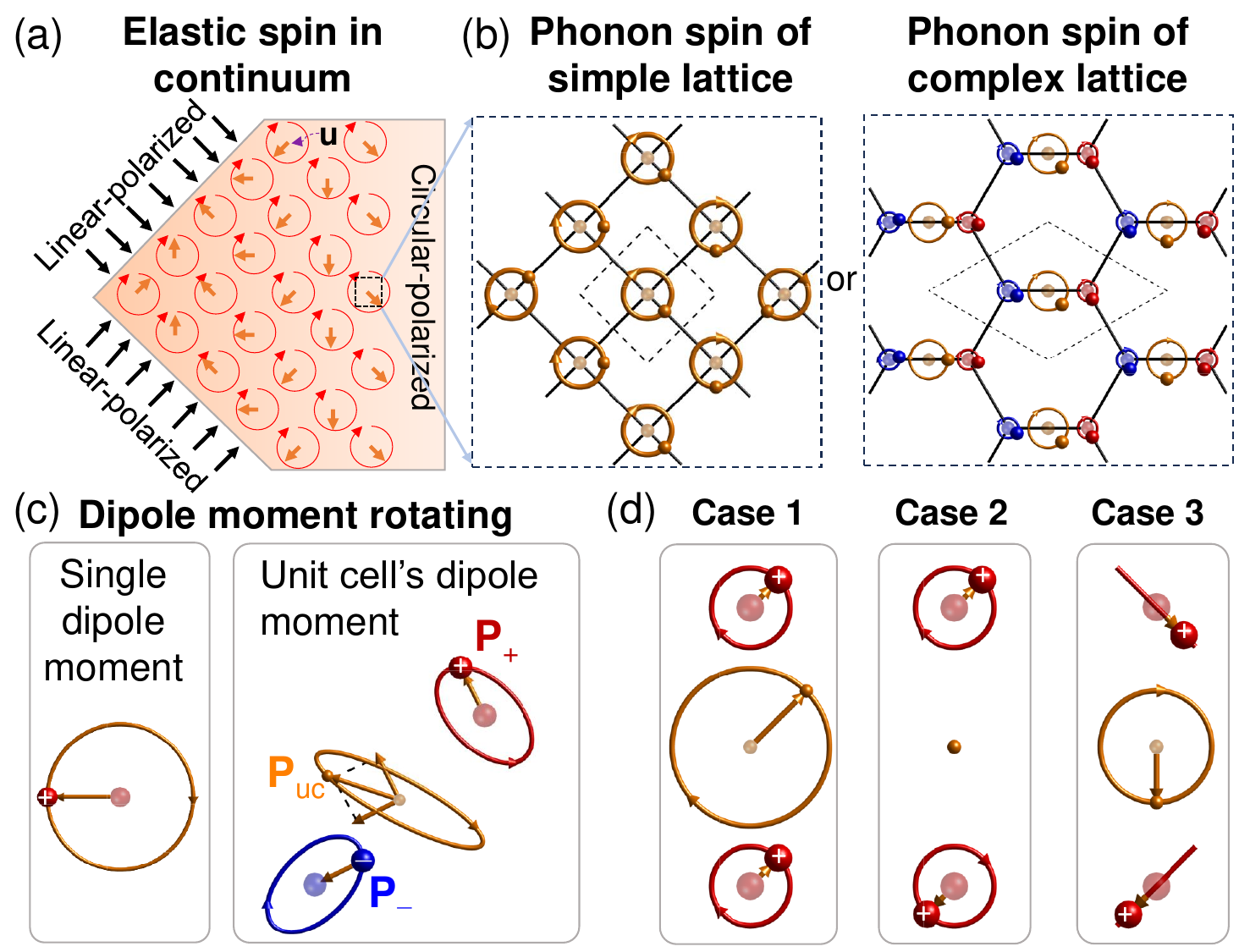}
\caption{Schematic diagram of elastic spin, phonon spin and dipole moment rotating (DMR). (a) Elastic spin in continuum. Composition of two out-of-phase linearly-polarized elastic wave will contribute to circularly-polarized elastic wave whose spin AM is nonzero. (b) Phonon spin in simple and complex lattices. The unit cell is marked by dashed parallelogram and the rotation of mass centers are shown by orange trajectories. The rotation of atoms are shown by red and blue trajectories in complex lattice at right panel. (c) Schematic illustration of DMR. Dipole moments shown by yellow arrows are caused by the positive- ($+$) or negative- ($-$) charged atoms' displacements from equilibrium position shown as red and blue arrows. The left panel shows the rotation of single dipole moment like the case of simple lattice and right panel shows the rotation of unit cell's DMR $\bm P_{\textrm{uc}}$ composited by every single dipole moment ($\bm P_\pm$) like the case of complex lattice. (d) Three cases of DMR in the diatomic unit cell. Case 1 shows nonzero DMR caused by two dipole moments rotating with the same phase. Case 2 shows DMR vanishes when two dipole moments rotating with phase difference $\pi$. Case 3 shows two linear motions of dipole moments with phase difference $\pi/2$ can contribute to nonvanish DMR.}
\label{sche}
\end{figure}

\textit{From elastic spin to phonon spin.} In the classical elastic field theory, the density of spin AM of a monotonic elastic wave can be calculated by ${\bm{s}}=\frac{\rho \omega}{2}\text{Im}(\bm{u}^*\times \bm{u})$ \cite{ElasticSpin2018PNAS} with the mass density $\rho$, angular frequency $\omega$ and displacement of mass element $\bm{u}$, which represents the rotation of the medium element's displaced polarization, as shown in Fig.~\ref{sche}(a). It can be expressed by a quantum-like matrix representation ${\bm{s}}= \langle \bs{u} |\hat{\bm{s}} | \bs{u} \rangle,$ with $\hat{\bm{s}}$ being spin-1 operator 
\begin{equation*}
    \hat{s}_x= 
    \begin{pmatrix}
        0  &  0 &  0\\
        0  &  0 & -i\\
        0  &  i &  0
    \end{pmatrix}
    , \hat{s}_y= 
    \begin{pmatrix}
        0  &  0  & i\\
        0  &  0  & 0\\
        -i &  0  & 0
    \end{pmatrix}
    , \hat{s}_z= 
    \begin{pmatrix}
        0  & -i & 0\\
        i  &  0 & 0\\
        0  &  0 & 0
    \end{pmatrix},
\end{equation*}
and $|\bs{u}\rangle = \sqrt{\frac{\rho \omega}{2}} \left( u_x, u_y, u_z \right)^T$ ($T$ stands for matrix transposition) being the normalized state vector of displacement \cite{PS2022CPL}. The elastic spin describes the local circular motions of mass elements. For simple lattices, like the isotropic cubic lattice or square lattice which contains only one atom within a unit cell, the displacement field vector $\bs{u}$ can be replaced by the atomic displacement \cite{1962PSLevine}, as shown in the left panel of Fig.~\ref{sche}(b). However, if there are more than one atom in the unit cell as for complex lattice, we should regard the unit cell, rather than the atom, as the basic element, which is exemplified with hexagonal lattice in the right panel of Fig.~\ref{sche}(b).

Generally, the displacement of mass center of unit cell is $\bm{u}_{\textrm{C}}=\sum_i \bar{M}_i \bm{u}_i$ with $\bar{M}_i$ being the reduced mass $\bar{M}_i=M_i/M_\textrm{C}$, $M_\textrm{C}=\sum_i M_i$ and $\bs{u}_i$ representing atomic displacement. For a single phonon mode with angular frequency $\omega$, the spin AM of unit cell is related to the rotating displacement polarization of mass center, calculated by $M_\textrm{C} \bm{u}_{\textrm{C}} \times \dot{\bm{u}}_{\textrm{C}}$ and its time-averaged value for unit cell including $n$ atoms is
\begin{equation}
    \bm{S}=\langle \bs{u}_{\textrm{uc}} |\hat{\bm{S}} | \bs{u}_{\textrm{uc}} \rangle=\bm{S}^{\textrm{local}}+\bm{S}^{\textrm{nonlocal}},
    \label{AMMC}
\end{equation}
where ${\bm{\hat{S}}}=\hat{M} \otimes \hat{\bm{s}}$ with $\hat{M}$'s matrix elements $M_{ij} = \bar{M}_i \bar{M}_j$ and $|\bs{u}_{\textrm{uc}}\rangle = \sqrt{\frac{M_{\textrm{C}}\omega}{2}} (\bs{u}_1, \bs{u}_2, \cdots, \bs{u}_n)^T$ is the displacement vector of the unit cell. As long as there are more than one atom ($n>1$) in the unit cell, both diagonal and non-diagonal elements of $\hat{M}$ will exist so that the angular momentum $\bm{S}$ can be divided into two parts: a local term  $\bm{S}^{\textrm{local}}=\sum_{i=1}^n M_{ii}\langle \bs{u}_{i} |\hat{\bm{s}} | \bs{u}_{i} \rangle$, and a nonlocal term $\bm{S}^{\textrm{nonlocal}}=\sum_{i\neq j}^n M_{ij}\langle \bs{u}_{i} |\hat{\bm{s}} | \bs{u}_{j} \rangle$. Specifically, the diagonal elements corresponding to local term describe the rotation of each atom referring to equilibrium position; while the off-diagonal elements describe the relative rotation of atoms referring to the other atoms which corresponds to a kind of nonlocal crossing interference effect. 
For phonons, which are coherent vibrational modes that naturally bring phase interference among many particles, this $\bm S^{\textrm{nonlocal}}$ will definitely make the collective interference phonon spin distinct from the previously used $\bm S^{\textrm{local}}$,  as we will demonstrate in detail in what follows.

\textit{Circularly dichroism and dipole moment rotating (DMR).}
Generally, phonon spin is able to couple with spin of other (quasi) particles, which gives rise to AM transfer processes. In the context of phonon-photon interaction, CP phonon modes can couple with external CP electromagnetic fields \cite{ChiralPhonon2018Science, ChiralPhonon2023NP, ChiralPhonon2023Nature} resulting in CP Raman scatterings \cite{Zhang2023Aug, Che2025May}. We are interested in the CP infrared absorption, i.e. phonon-induced infrared circular dichroism (ICD). Traditionally, the ICD relies on the coexistence of electric and magnetic dipole moments together \cite{Keiderling2020Apr,Nafie2011,Stephens1985Feb,Stephens1987Mar} but recent studies have shown that magnetic dipoles are not necessary for ICD \cite{Beaulieu2018May} and electric dipoles alone can give rise to giant circular polarized light emission \cite{Wan2023Feb}. 

Here, we investigate phonon-induced ICD in nonmagnetic solids based on the infrared Hamiltonian as \cite{Peter2010}
\begin{equation}\label{H_infrared}
H = \sum_{\ell i} Q_i \bs{u}_{\ell i} \cdot \bs{E}_{\ell i}, 
\end{equation}
where $\bs{u}_{\ell i}$ represents the ionic displacement of $i$-th atom belonging to $\ell$-th unit cell, $\bs{E}_{\ell i}$ denotes the optical electric field on that atom, and $Q_i$ refers to the (effective) ionic charge. Here, the electric dipole approximation for the light-matter interaction Hamiltonian is used and it is equivalent to the formulism by using vector potential $\bs{A}$ (see the comparison between two different formulisms at the end of section S1 of SM \cite{SM}). According to the Fermi golden rule, the transition rates of this infrared phonon-photon process are determined by
\begin{equation}\label{I_fermi_golden}
    I=\frac{2 \pi}{\hbar} \sum_{i,f} p_i |\langle \Phi_f | \hat{H} | \Phi_i \rangle|^2 \delta(E_f - E_i \pm \hbar \omega),
\end{equation}
where $|\Phi_{i/f}\rangle$ refers to initial/final phonon eigenstates under occupation number representation and $p_i$ is the probability of initial states. Energy conservation is ensured by delta function of initial (final) energy $E_i$ ($E_f$) and phonon/photon energy $\hbar \omega$. Here, we consider a CP light $\bm{E}^{R/L}_{\ell i}=\frac{E_0}{\sqrt{2}}(\bs{e}_1\pm i \bm{e}_2) \exp(i\bs{q} \cdot \bs{r}_{\ell i})$ with $\bs{r}_{\ell i} = \bs{R}_\ell + \bs{\tau}_i$ being the ionic position vector. $\bs{R}_\ell$ is the lattice vector and $\bs{\tau}_i$ is the coordinate of $i$-th sublattice within the unit cell. $\bm{e}_{\bm{q}} \equiv \bs{q}/|\bs{q}| =\bm{e}_1 \times \bm{e}_2$ is the direction vector of incident light. $L$ and $R$ refer to the left- and right-handedness of the light, respectively. By substituting the second quantization form of phonon $\bm{u}_{\ell i}  = \sum_{n\bm{k}} \sqrt{\frac{\hbar}{2N M_i\omega_{n\bm{k}}}} \bm{\varepsilon}^i_{n\bm{k}} \hat{a}_{n\bm{k}} e^{i(\bm{k}\cdot \bm{r}_{\ell i} - \omega_{n\bm{k}}t)}+\textrm{H.c.}$ into Eqs. \eqref{H_infrared} and \eqref{I_fermi_golden}, we can get the difference of transition rates between left and right CP infrared light $\Delta I(\bs{q},\omega) =I_L-I_R$, i.e., the ICD, as (see details in section S1 of supplemental material (SM)~\cite{SM}) 
\begin{equation}
    \frac{\Delta I(\bs{q},\omega)}{N}
        =\sum_{n} \frac{\pi |E_0|^2}{\hbar \omega} \delta(\omega-\omega_{n\bm{q}}) |f(\pm\hbar \omega_{n\bm{q}})| \bm{e}_{\bm{q}} \cdot \bm{\mathcal{R}}_{n\bm{q}},
    \label{CD}
\end{equation}
where $N$ is the number of unit cell, $\omega_{n\bm{q}}$ is phonon (angular) frequency of the $n$-th branch at wave vector $\bm{q}$ of the incident light; $f(\hbar \omega)=(e^{\hbar \omega/k_B T}-1)^{-1}$ is the Bose-Einstein distribution function with ``$\pm$'' in Eq. \eqref{CD} corresponding to the photon-absorption-phonon-emission ($+$) or photon-emission-phonon-absorption ($-$) processes, respectively; $\bm{\mathcal{R}}_{n\bm{q}}$ describes the dipole moment rotating (DMR) of the unit cell under the phonon mode $(n,\bm q)$, reading as
\begin{equation}
    \bm{\mathcal{R}}_{n\bm{q}}= \sum_{i,j} {\text{Im}} (Q_i \bm{u}_{n\bm{q}}^{i *}\times Q_j \bm{u}_{n\bm{q}}^j) = \textrm{Im} (\bm{P}^*_{\textrm{uc}} \times \bm{P}_{\textrm{uc}}),
    \label{DMR}
\end{equation}
where the indices $i$, $j$ are summed over the atom index within the unit cell and $\bm{u}_{n\bm{q}}^i=\bm{\varepsilon}_{n\bm{q}}^i /\sqrt{M_i}$ describes the displacement of $i$-th atom corresponding to the eigenvector $\bs{\varepsilon}_{n\bs{q}}$. $\bm{P}_{\textrm{uc}}=\sum_i Q_i \bm{u}_i$ is the electric charge dipole moment of the unit cell. As such, the DMR $\bs{\mathcal{R}}_{n\bs{q}}$ is proportional to the time-average of $\bs{P}_{\textrm{uc}} \times \dot{\bs{P}}_{\textrm{uc}}$, which clearly describes the rotation of the unit cell's dipole moment $\bs{P}_{\textrm{uc}}$.

We give a schematic illustration of DMR in Fig.~\ref{sche}(c) with simple lattice (left panel) and complex lattice (right panel).
It can also be expressed in matrix form as
\begin{equation}
    \bm{\mathcal{R}}=\langle \bs{u}_{\textrm{uc}} |\hat{\bs{\mathcal{R}}} | \bs{u}_{\textrm{uc}} \rangle 
    =\underbrace{\sum^n_{i=1} Q_{ii}\langle \bs{u}_{i}|\hat{\bm{s}}|\bs{u}_{i} \rangle}_{\bm{\mathcal{R}}^{\textrm{local}}}+\underbrace{\sum^n_{i\neq j} Q_{ij}\langle \bs{u}_{i}|\hat{\bm{s}}|\bs{u}_{j} \rangle}_{\bm{\mathcal{R}}^{\textrm{nonlocal}}},
    \label{DMRMat}
\end{equation}
where $\hat{\bm{\mathcal{R}}}=  \hat{Q} \otimes \bm{\hat{s}}$, with matrix element of $\hat{Q}$ given by $Q_{ij}=\frac{2}{M_{\textrm{C}}\omega}Q_i Q_j$ and $\bm{\hat{s}}$ being the spin-1 operator. Note that $\bm{\mathcal{R}}$ has a similar structure as $\bm{S}$ in Eq.~\eqref{AMMC}. Obviously, the nonvanishing off-diagonal elements $Q_{ij}\langle \bs{u}_{i}|\hat{\bm{s}}|\bs{u}_{j} \rangle$ indicates the similar nonlocal crossing interference effect between different sublattices as the situation of unit cell's spin AM, i.e., the term $\bm{S}^{\textrm{nonlocal}}$ in Eq.~\eqref{AMMC}. From the theoretical derivation, we clearly see that not only the local rotation of individual atoms, but also the nonlocal interference among many particles should be considered in the spin AM transfer between phonon-photon process, which however has been overlooked in many previous studies. It should be noted that the ICD caused by DMR is conceptually different from the conventional vibrational circular dichroism (VCD) of molecules caused by optical rotational strength \cite{Stephens1985Feb, Stephens1987Mar, Nafie2011, Keiderling2020Apr} in three aspects. First, their mathematical expressions are totally different [see Eq. (S1.1) in SM \cite{SM}]. Second, they have different physics that the DMR here originates from the electrical dipole interactions while the VCD requires interactions between electrical dipole and magnetic dipole. Finally, the applicable physical systems are different: the DMR introduced here is applied to crystalline solids while VCD is typically applied to molecules, so that DMR applies to phonon bands with finite $\bm q$ in BZ while VCD can be regarded as specialized at $\Gamma$ point.

For a deeper understanding of the physical insight of DMR in complex lattice, we exemplify three cases in diatomic lattice in Fig.~\ref{sche}(d) to make things more comprehensible. The first case shows that two atoms with identical charge rotate with the same phase contributing to a nonzero unit cell's dipole moment and DMR. This is an ordinary case that nonlocal effect vanishes since there is no relative rotation between two atoms. By tuning the phase difference between two atoms, the contribution from local rotation keeps the same while the nonlocal crossing interference effect becomes significant. The total DMR will vanish when two atoms' motions are in anti-phase, shown as the second case in Fig.~\ref{sche}(d), since the unit cell's dipole moment keeps still. In other word, the nonlocal collective interference effect exactly cancels the local atom rotation in case 2. The third case shows that even two linear motions of atoms can contribute to nonzero DMR, where only nonlocal crossing interference effect exists, in the absence of local atom rotation. More vivid schematic illustrations of two-atom model showing the distinct physical pictures of $\bm{S}^{\textrm{nonlocal}}$ and $\bm{S}^{\textrm{local}}$ are given in Fig. S1 of \cite{SM} and video in \cite{Data}.

\begin{figure}[tbp]
\centering	
\includegraphics[width=0.48\textwidth]{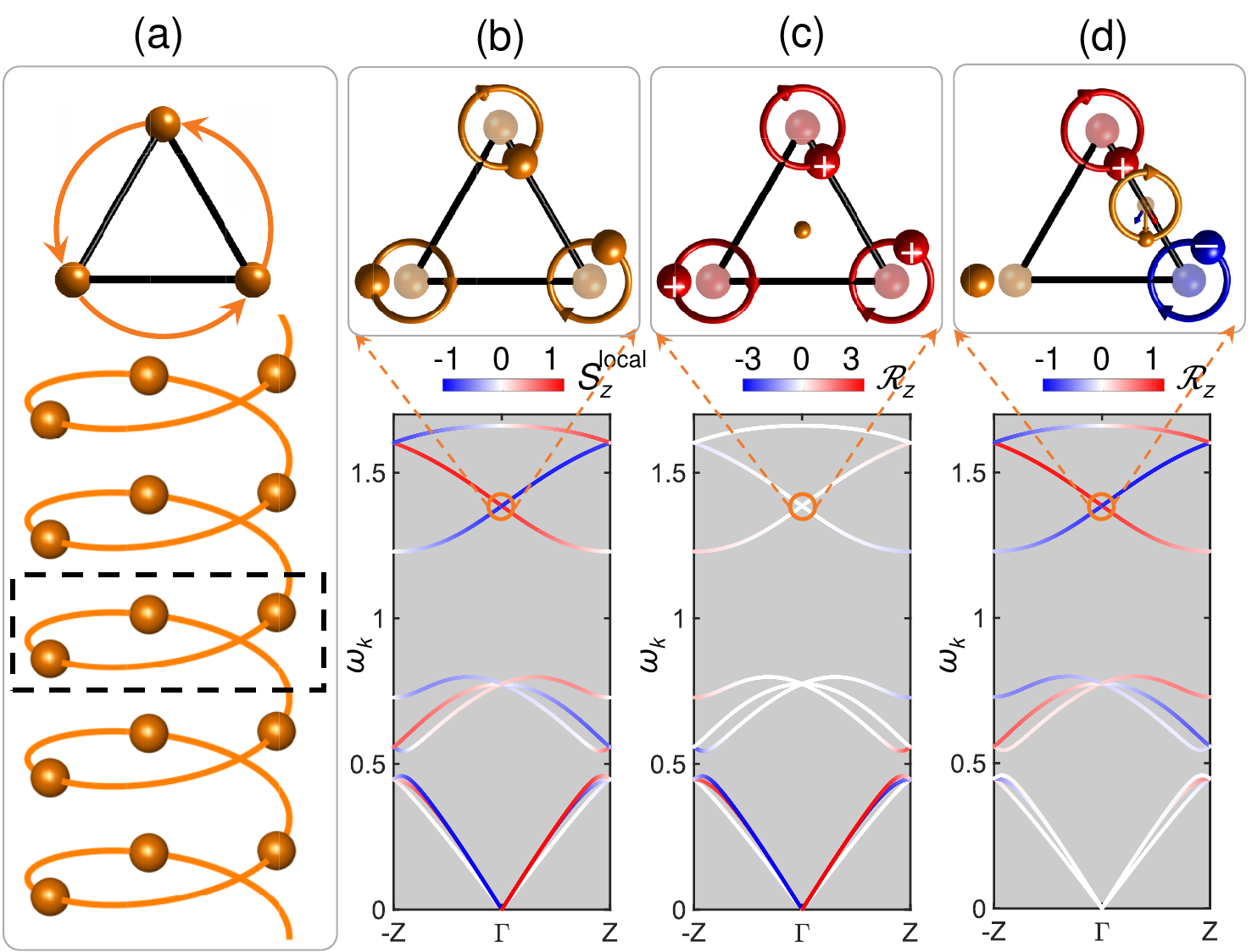}
\caption{DMR mapped on phonon dispersion relation of 3-fold helical chain. (a) Schematic diagram of three-fold screw helical chain. The unit cell is marked by black dotted box and the top view of this structure is shown above. (b) $S_z^{\textrm{local}}$ component of AM of atoms' local rotation mapped on the dispersion relation. The motion of atoms at $\Gamma$ point for one of the two-degenerate optical states is shown on the top where three atoms rotate clockwise. (c) $\mathcal{R}_z$ component of DMR mapped on the dispersion relation with charge distribution $[+1,+1,+1]$. The $\mathcal{R}_z$ mainly exists at acoustic branches while keeps zero at optical branches since atoms' motion at optical branches are out-of-phase resulting zero dipole moment of unit cell which is shown above by orange solid arrows. (d) $\mathcal{R}_z$ component of DMR mapped on the dispersion relation with charge distribution $[+1,-1,0]$. Nonzero DMRs arise at optical branches due to the rotating of unit cell's dipole moment. }
\label{fig2}
\end{figure}

\begin{figure}[tbp]
\centering
\includegraphics[width=0.48\textwidth]{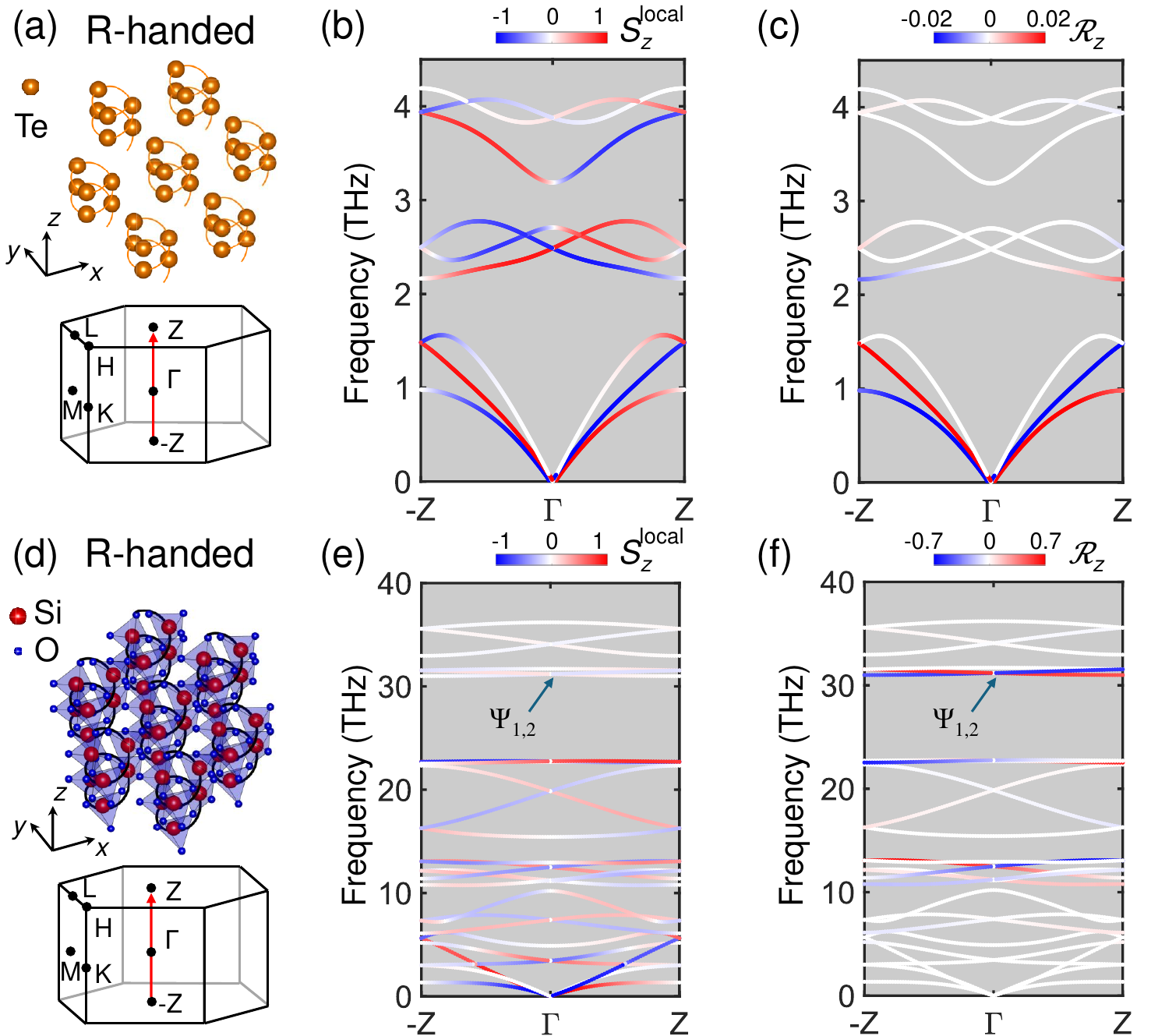}
\caption{Comparison between the AM of local atom rotation $S_z^{\textrm{local}}$ and the collective interference DMR $\mathcal{R}_z$ of real chiral materials possessing different charge distribution. (a),(d) Atomic structure of right-handed tellurium and $\alpha$-quartz, respectively. (b)-(c) Phonon dispersion relation of tellurium together with band resolved $S^{\textrm{local}}_z$ and $\mathcal{R}_z$, respectively. (e)-(f) Same as (b)-(c) but for $\alpha$-quartz. The results clearly demonstrate that the DMR, as a manifestation of collective interference phonon spin in ICD, is for  a coherent vibrational mode involving collective motion of many particles, so that is distinct from the local rotation of individual atom. See details of the calculation methods in section S2 of SM \cite{SM}. }
\label{fig3}
\end{figure}

\textit{A chiral model study of DMR.} Since the DMR is a pseudo vector, it vanishes throughout the whole BZ in nonmagnetic centrosymmetric materials because of the combined inversion and time-reversal symmetries. To support nonzero DMR in nonmagnetic materials, inversion symmetry must be broken, which is typical in chiral lattices. As an example, we investigate DMR in a helical chain model with 3-fold skew rotation [Fig. \ref{fig2}(a)]. 
Figure \ref{fig2}(b) shows the hourglass-like phonon dispersion relation (lower panel) with the mapping of the diagonal part of phonon spin, i.e., $\bs{S}^{\textrm{local}}$. Red and blue colors indicate the positive and negative values of $S^{\textrm{local}}_z$. $S^{\textrm{local}}_{x,y}$ vanish because of the 3-fold skew rotation symmetry. Clearly, all branches support nonzero $S^{\textrm{local}}_z$. Figure  \ref{fig2}(c) shows the DMR of uniform charge, or equivalently, the total phonon spin $\bs{S}$, which includes the nonlocal contribution from the many-particle-interference. Because the DMR is related to the AM of the unit cell, $\mathcal R_z$ mainly distributes on the transverse acoustic branches whereas its value on optical branches is relatively small. However, when the atomic charge distribution is nonuniform, the DMR of the optical branches can be nonzero, see Fig. \ref{fig2}(d).

\textit{DMR and ICD in realistic chiral materials.} 
We further implement first-principles calculations of DMR for two real chiral materials, tellurium (Te) and $\alpha$-quartz (SiO$_2$). The structures of right-handed Te and SiO$_2$ are shown in Figs.~\ref{fig3}(a) and (d), respectively. The ionic charge distributions of these two chiral crystals represent the typical examples of uniform and nonuniform charge distributions (c.f. Figs. \ref{fig2}(c)-(d)). By mapping the local rotation $S_z^{\textrm{local}}$ and the collective interference DMR $\mathcal{R}_z$ on the phonon dispersion relation of these two materials in Figs.~\ref{fig3}(b)-(c) and (e)-(f), we find: (i) for the uniform charge distributed Te, $\mathcal{R}_z$ almost vanishes at the optical branches while $S_z^{\textrm{local}}$ are nonzero at both acoustic and optical branches, which meets the situation of Fig.~\ref{fig2}(c); (ii) for nonuniform charge distributed $\alpha$-quartz that meets the situation of Fig.~\ref{fig2}(d), nonvanishing $\mathcal{R}_z$ emerges at the optical branches, whose behaviors are clearly distinct from $S_z^{\textrm{local}}$.

\begin{figure}[tbp]
\centering
\includegraphics[width=0.48\textwidth]{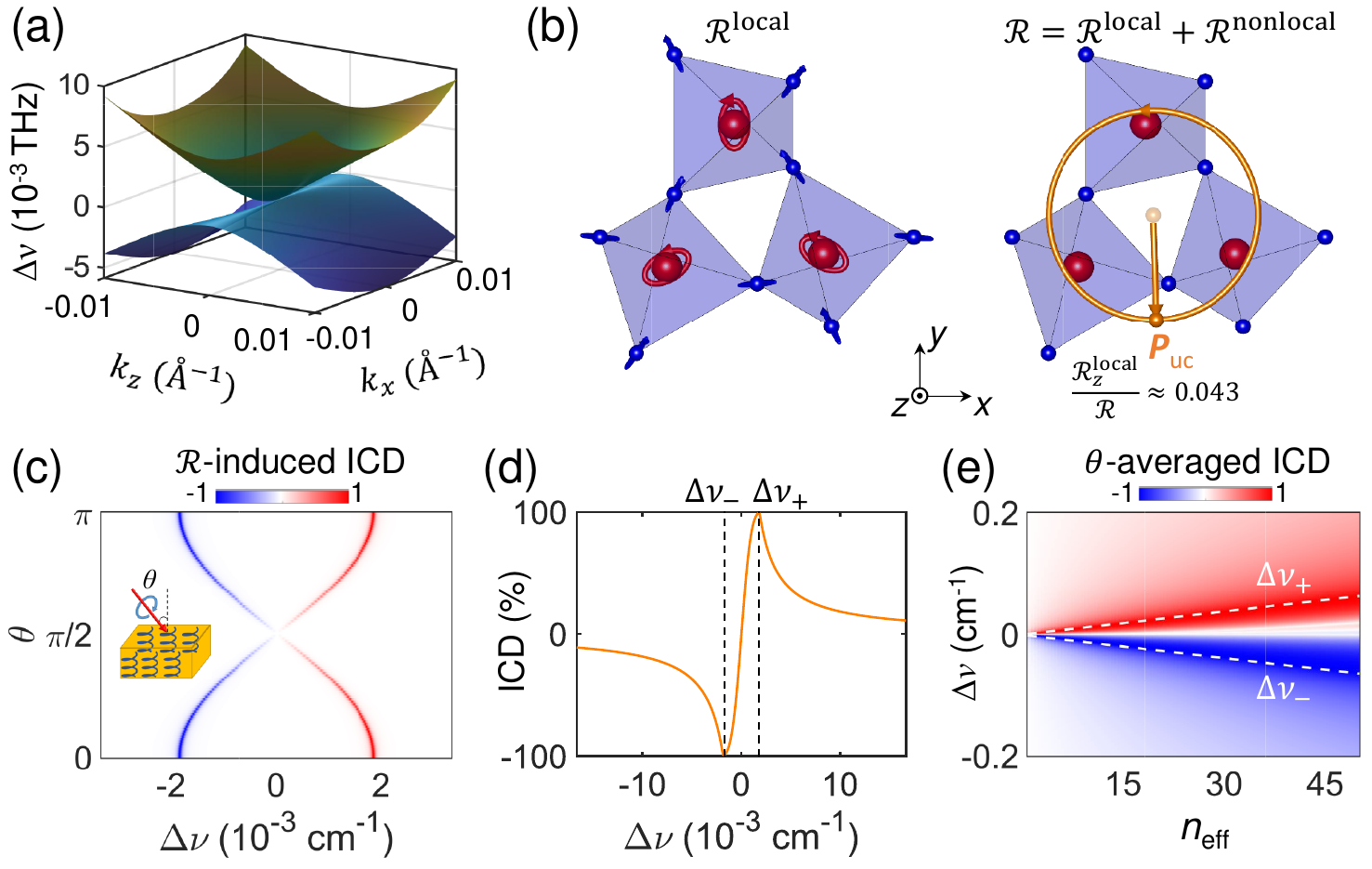}
\caption{Experimental measurable signature for the DMR, as a manifestation of collective interference phonon spin in ICD. (a) Phonon dispersion for the Weyl phonon [$\Psi_{1,2}$ states in Figs. \ref{fig3}(e)-(f)] around $\Gamma$. $\Delta \nu$ denotes the light frequency difference referenced to the frequency of Weyl phonon. (b) $\mathcal{R}^{\textrm{local}}$ (left panel) and total DMR $\mathcal{R} = \mathcal{R}^{\textrm{local}} + \mathcal{R}^{\textrm{nonlocal}}$ (right panel) corresponding to state $\Psi_1$. The degenerate partner state $\Psi_2$ has opposite $\mathcal{R}^{\textrm{local}}$ and $\mathcal{R}$ as $\Psi_1$. (c) Calculated spectrum of $\bs{\mathcal{R}}$-induced ICD for right-handed $\alpha$-quartz as a function of incident angle $\theta$ and frequency $\Delta \nu$. The incident angle is defined as the angle between the incident direction ($\bs{e_q}$) and the skew rotation symmetry axis as shown in the inset. (d) $\bs{\mathcal{R}}$-induced ICD averaged over incident angle $\theta$. $\Delta \nu_+$ ($\Delta \nu_-$) refers to the frequency of positive (negative) peak. (e) $\theta$-averaged ICD spectrum as function of effective refractive index $n_{\textrm{eff}}$.}
\label{fig4}
\end{figure}

From Eq.~\eqref{CD}, the ICD of chiral crystal will be a function of light wave vector $\bs{q}$ that determines the angle $\theta$ of incident light. Here we take the two degenerate states $\Psi_{1,2}$ [shown in Fig.~\ref{fig3}(f)] at frequency 31.19 THz of right-handed $\alpha$-quartz as an example, which can be described by a quadratic Weyl model with Chern number $\pm2$ \cite{Wang2020Mar, Lange2024Mar, SM}, with the enlarged 3D phonon dispersion shown in Figure \ref{fig4}(a). Besides topological nature, the $\Psi_{1,2}$ states also exhibit significant DMR ($\bs{\mathcal{R}}$) and thus $\bs{\mathcal{R}}$-induced ICD. Figure \ref{fig4}(b) shows the enlarged $\mathcal{R}^{\textrm{local}}$ corresponding to phonon eigenvector $\Psi_1$ and the corresponding total DMR $\mathcal{R}$. The Si (O) ions with positive (negative) charges vibrate conter-clockwisely (clockwisely). 
The charge dipoles of Si and O ions can add up to a large $\bs{P}_{\textrm{uc}}$, whose rotating thus forms large DMR $\mathcal{R}$ of the unit cell. Our first-principles calculation shows that the collective interference effects dominate in the DMR and more than 95\% of the DMR comes from nonlocal interference term $\mathcal{R}^{\textrm{nonlocal}}$.

Figure \ref{fig4}(c) shows the angle-dependent $\bs{\mathcal{R}}$-induced ICD spectrum for single-crystalline right-handed $\alpha$-quartz. The ICD signal gets maximized when the incident direction is parallel to the skew rotation axis ($\theta = 0, \pi$). $\mathcal{R}_x$ and $\mathcal{R}_y$ vanish as a result of skew rotational symmetry (see Fig. S4 in \cite{SM}). When the chiral $\alpha$-quartz is polycrystalline powder with random orientations, we average over the incident angle, and the absorption curves for $\sigma_\pm$ exhibit different peaks on both sides of the frequency of Weyl phonon [see Fig.~\ref{fig4}(d)].
The two-peak interval can be enlarged by slowing down the group velocity of infrared light. This could be achieved by utilizing the infrared surface plasmon polariton (SPP) at the interface between chiral $\alpha$-quartz and metallic metasurface, whose speed can be slowed down with arbitrarily designed high value of refractive index \cite{Shen2005May}. Figure \ref{fig4}(e) shows the angle-averaged ICD spectrum as function of effective refractive index. The separation of positive and negative ICD signals can easily reach 0.1 cm$^{-1}$ which is more than accessible for current experimental techniques \cite{Keiderling2018Sep}. Similar SPP-enhanced spectral separation techniques have been validated in recent experiments \cite{Vicentini2025Dec}. 

In conclusion, we have demonstrated the significant role of collective motion of many particles in phonon spin of complex lattice, where the nonlocal crossing interference exhibits measurable effects. 
In particular, we have formulated the CP infrared photon-phonon process by quantum perturbation theory and obtained the ICD specturm in terms of a physical quantity DMR arising from the rotating motion of unit cell's dipole moment. This DMR is strongly correlated with the nonlocal crossing interference phonon spin and has an pivotal impact on the measurable ICD spectrum in complex lattices.
Results have been exemplified in both theoretical chiral structures and realistic chiral materials.

Our study uncovers a deep insight of collective interference phonon spin in real materials with complex lattices. Although we focus on infrared absorption process at present, similar effects is expected to apply for the Raman scattering in chiral materials. Moreover, besides the spin AM transfer in photon-phonon process, the spin degree of freedom of phonon is able to interact with the spin of many other (quasi) particles as well, through such as magnon-phonon interaction~\cite{MagnonPS2018NP,MagnonPS2018PRL,MagnonPS2020PRL,Magnon2022Nature} and electron-phonon interaction~\cite{ElectronPS2020PRR, ElectronPS2023PRR, ElectronPS2024PRLE, ElectronPS2024PRLT}, which would exhibit more plentiful phenomena. The spin phononics with coupling to multiple degrees of freedom, such as electron spin, photon spin, magnon spin, and so on, can provide opportunity to develop solid-based devices for more promising functionalities.

This work is supported by the National Key R\&D Program of China (No. 2023YFA1406900 and No. 2022YFA1404400), the Innovation Program for Quamtum Science and Technology (No. 2023ZD0300500), the National Natural Science Fundation of China (No. 12404279)， the Natural Science Foundation of Shanghai (No. 23ZR1481200), the Program of Shanghai Academic Research Leader (No. 23XD1423800), Shanghai Pujiang Program (No. 23PJ1413000), and the Fundamental Research Funds for the Central Universities. 

\textit{Data availabity}---The data that support the findings of this article are openly avaiable \cite{Data}.


\providecommand{\noopsort}[1]{}\providecommand{\singleletter}[1]{#1}%

\end{document}